\documentclass[conference,letterpaper]{IEEEtran}
\IEEEoverridecommandlockouts

\usepackage[utf8]{inputenc} 
\usepackage[T1]{fontenc}
\usepackage{url}
\usepackage{ifthen}
\usepackage{cite}
\usepackage[cmex10]{amsmath} 
\usepackage{amsfonts, amssymb}
\usepackage{mathtools}
\usepackage{manfnt}
\usepackage{contour}
\usepackage[inline]{enumitem}
\usepackage{booktabs}
\usepackage{xfrac}
\usepackage{algorithmic}
\usepackage{url}
\usepackage{graphicx}
\usepackage{textcomp}
\usepackage{xcolor}
\usepackage{standalone}
\usepackage{xspace}
\usepackage{etoolbox}
\ifCLASSOPTIONcompsoc
 \usepackage[caption=false,font=normalsize,labelfont=sf,textfont=sf]{subfig}
\else
 \usepackage[caption=false,font=footnotesize]{subfig}
\fi

\usepackage{tikz}
\usetikzlibrary{%
    arrows.meta,%
    calc,%
    colorbrewer,%
    decorations.pathreplacing,%
    matrix,%
    math,%
    patterns,%
    positioning,%
    shadows,%
}

\usepackage{comment}
\usepackage[capitalize]{cleveref}
\usepackage{microtype}
\usepackage{thmtools}
\usepackage{thm-restate}

\crefformat{section}{\S#2#1#3}
\crefmultiformat{section}{\S\S#2#1#3}{ and~#2#1#3}{, #2#1#3}{, and~#2#1#3}
\crefname{example}{Ex.}{Exs.}

\DeclarePairedDelimiter{\parens}{\lparen}{\rparen}

\newcommand*{\Initial}{I\mskip-2mu}
\newcommand*{\Final}{F\mskip-2mu}
\newcommand*{\In}{n^{\Initial}}
\newcommand*{\Fn}{n^{\Final}}
\newcommand*{\Ik}{k^{\Initial}}
\newcommand*{\Fk}{k^{\Final}}
\newcommand*{\Ip}{\V{P^{\Initial}}}
\newcommand*{\Fp}{\V{P^{\Final}}}
\newcommand*{\IPart}{{\mathcal{P}}^\Initial}
\newcommand*{\FPart}{{\mathcal{P}}^\Final}

\newcommand*{\Ir}{r^{\Initial}}
\newcommand*{\Fr}{r^{\Final}}

\newcommand*{\Il}{\lambda^{\Initial}}
\newcommand*{\Fl}{\lambda^{\Final}}

\newcommand*{\V}[1]{\mathbf{#1}}

\newcommand*{\Code}{\mathcal{C}}

\newcommand*{\ICode}{\Code^{I}}
\newcommand*{\FCode}{\Code^{F}}

\definecolor{color-r}{HTML}{d95f02}

\newcommand{\F}{\mathbb{F}}
\renewcommand{\Pr}{\mathop{\bf Pr\/}} 
\newcommand{\lam}{\lambda}
\newcommand{\calH}{\mathcal{H}}
\newcommand{\calM}{\mathcal{M}}

\newtheorem{corollary}{Corollary}
\newtheorem{lemma}{Lemma}
\newtheorem{theorem}{Theorem}

\newtheorem{definition}{Definition}
\newcommand\xqed[1]{%
  \leavevmode\unskip\penalty9999 \hbox{}\nobreak\hfill
  \quad\hbox{#1}}
\newcommand{\triqed}{\xqed{\mantriangleright}}
\AtEndEnvironment{example}{\triqed}

\begin{document}
\title{On Low Field Size Constructions of \\ Access-Optimal Convertible Codes\\
\thanks{
This work was funded in part by an NSF CAREER award (CAREER-1943409), a Sloan Fellowship and a VMware Systems Research Award.
}} 

\author{%
 \IEEEauthorblockN{Saransh Chopra, Francisco Maturana, and K. V. Rashmi}
 \IEEEauthorblockA{Computer Science Department\\
                   Carnegie Mellon University\\
                   Pittsburgh, PA, USA\\
                   Email: \{saranshc,fmaturan,rvinayak\}@andrew.cmu.edu}
}

\maketitle

\begin{abstract}
   Most large-scale storage systems employ erasure coding to provide resilience against disk failures. Recent work has shown that tuning this redundancy to changes in disk failure rates leads to substantial storage savings. This process requires \textit{code conversion}, wherein data encoded using an $[\In,\Ik]$ initial code has to be transformed into data encoded using an $[\Fn,\Fk]$ final code, a resource-intensive operation. \emph{Convertible codes} are a class of codes that enable efficient code conversion while maintaining other desirable properties. In this paper, we focus on the \emph{access cost} of conversion (total number of code symbols accessed in the conversion process) and on an important subclass of conversions known as the merge regime (combining multiple initial codewords into a single final codeword). 
   
   In this setting, explicit constructions are known for systematic access-optimal Maximum Distance Separable (MDS) convertible codes for all parameters in the merge regime. However, the existing construction for a key subset of these parameters, which makes use of Vandermonde parity matrices, requires a large field size making it unsuitable for practical applications. In this paper, we provide (1) sharper bounds on the minimum field size requirement for such codes, and (2) explicit constructions for low field sizes for several parameter ranges. In doing so, we provide a proof of super-regularity of specially designed classes of Vandermonde matrices that could be of independent interest.
\end{abstract}

\section{Introduction}
\label{sec:introduction}

    Erasure codes are used widely in modern large scale distributed storage systems as a means to mitigate data loss in the event of disk failures. In this context, erasure coding involves dividing data into groups of $k$ chunks that are each encoded into stripes of $n$ chunks using an $[n,k]$ erasure code. These encoded chunks are then stored across $n$ distinct storage nodes in the system. The code parameters $n$ and $k$ determine the amount of redundancy added to the system and the degree of durability guaranteed.

    There are various classes of codes that are commonly used in real-world systems. For example, \emph{systematic} codes are those in which the original message symbols are embedded among the code symbols. This is highly desirable in practice as in the event that there are no observed disk failures, there is no decoding process needed to recover the original data. Systematic codes with \emph{Vandermonde parity matrices} (see \cref{sec:MDS_Systematic_Vandermonde_Codes}) are even more advantageous as there are known efficient algorithms utilizing Fast Fourier Transform (FFT) for computing the product between vectors and Vandermonde matrices \cite{Gohberg_Olshevsky_1994, lacan_fimes_2004}, speeding up the encoding process. This attribute is becoming increasingly important given the recent trend to use wider (high $k$) and longer (high $n$) erasure codes \cite{Hu_et_al_2021_wide,kadekodi_et_al_2023_wide}. Additionally, \emph{Maximum Distance Separable (MDS)} codes are a subset of erasure codes that require the least amount of additional storage in order to meet a specific failure tolerance goal. An $[n,k]$ MDS code can tolerate loss of any $n-k$ out of the $n$ code symbols. In this paper, our interest is on systematic MDS codes with Vandermonde parity matrices. 
    
    Recent findings by Kadekodi et al. \cite{kadekodi2019cluster} reveal the dynamic variability in disk failure rates over time. Their research highlights the potential for meaningful savings in storage and associated operational expenses through tuning code parameters to observed failure rates. However, the resource overhead associated with the \emph{default approach} of re-encoding all of the data in order to modify $n$ and $k$ is prohibitively expensive~\cite{maturana2022convertible}. 

The \emph{code conversion} problem introduced in~\cite{maturana2022convertible} formalizes the problem of efficiently transforming data that has been encoded under an $[\In, \Ik]$ initial code $\ICode$ to its new representation under an $[\Fn, \Fk]$ final code $\FCode$.  One of the key measures of the cost of conversion is the \textit{access cost}, which represents the total number of  code symbols accessed (read/written) during conversion. \emph{Convertible codes}~\cite{maturana2022convertible} are a class of codes that enable efficient conversion while maintaining other desirable properties such as being MDS and systematic (more details in \cref{sec:convertible_codes}).

Among various types of conversions, the \emph{merge regime}, where $\Fk = \lam \Ik$ for any integer $\lam \ge 2$ (i.e., combining multiple initial codewords into a single final codeword), is the most important one. First, the merge regime requires the least resource utilization~\cite{maturana2020access} among all types of conversions and hence are a highly favorable choice for practical systems. Second, constructions for the merge regime are key building blocks for the constructions for codes in the \emph{general regime} which allows for any set of initial parameters and any set of final parameters~\cite{maturana2020access}. In this paper, our focus is on systematic MDS convertible codes in the merge regime.

   In \cite{maturana2022convertible}, the authors established lower bounds on the access cost of conversion and provided constructions of \emph{access-optimal} convertible codes for all parameters in the merge regime. Let us denote $\Ir := \In - \Ik$ and $\Fr := \Fn - \Fk${, (which correspond to the number of parity symbols in the initial and final codes if the codes are systematic). For cases where $\Ir > \Fr$ (i.e., when the initial configuration has more parities than the final configuration), the authors provide explicit constructions of systematic MDS access-optimal convertible codes over fields of size linear in $\Fn$. For cases where $\Ir < \Fr$ (i.e., when more parities are needed in the final configuration than in the initial), it has been shown~\cite{maturana2022convertible} that the access cost of conversion for MDS erasure codes is lower bounded by that of the default approach to decode and re-encode all of the data. As a consequence, it is not possible to realize any savings with specialized code constructions.

    However, in the case where $\Ir = \Fr$, the best-known construction requires a minimum field size of $p^D$ for any prime $p$ and some $D \in \Theta((\Fn)^3)$ \cite{maturana2022convertible}. This field size is far too high for efficient practical implementations. Most current instruction-set architectures are optimized to operate on bytes of data at a time. Utilizing erasure codes defined over larger field sizes can hamper the encoding/decoding speed. Hence most (if not all) practical implementations of storage codes use $\F_{256}$ (which translates each field symbol to a one-byte representation). Thus, the problem of constructing low-field-size access-optimal convertible codes remains open for the case $\Ir = \Fr$.

    In this paper, we study the setting of systematic MDS access-optimal convertible codes in the merge regime in the case where $\Ir = \Fr$. The best known construction of convertible codes in this setting is a systematic code with a very specific choice of super-regular  Vandermonde parity matrix with a singular degree of freedom \cite{maturana2022convertible} (as will be detailed in \cref{sec:MDS_Systematic_Vandermonde_Codes}). In \cref{Sec:fundamental_limits}, we improve on this construction by allowing more freedom in the choice of \emph{scalars} of the Vandermonde matrix. We then study the minimum field size $q^*(k,r)$ required for existence of the underlying $k \times r$ super-regular Vandermonde parity matrices of such codes. We provide two lower bounds on the minimum field size required, one applicable for codes over general prime power fields (\cref{thm:kr_bound}) and one for codes over fields of characteristic $2$ where $k > r$ (\cref{thm:2^r_bound}). For fields of characteristic $2$, the bound takes the form $q^*(k,r) \ge \Omega(2^{r})$. Additionally, we establish an upper bound $q^*(k,r) \le O(k^r)$ (\cref{thm:k^r_bound}), which in turn results in an improved upper bound $q \le O((\Fk)^{\Fr})$ on the field size required for the existence systematic MDS access-optimal convertible codes in the merge regime in the case where $\Ir = \Fr$.
    
    Furthermore, in \cref{Sec:low_field_size_constructions}, we provide the first explicit low-field-size constructions of convertible codes in this setting for several parameter ranges via constructing their corresponding super-regular Vandermonde parity matrices. The proposed construction makes use of field automorphisms in designing the Vandermonde matrices. For any general prime power field $\F_q$ where $q = p^w$, we find explicit constructions of $k \times 3$ super-regular Vandermonde matrices for all $k$ such that $k < w$ (\cref{thm:general_field_construction}).  This, in turn, gives us a construction of systematic MDS access-optimal convertible codes for all parameters in the merge regime such that $\Fr = \Ir \le 3$ and $\Fk < w$. For any finite field $\F_q$ where $q = 2^w$ (that is, characteristic 2), we present a stronger result covering a larger range of $k$ by showing that the same proposed construction is super-regular for all $k$ such that $k < q$ (\cref{thm:binary_field_construction}). 
    
    These results are also of independent interest beyond the setting considered in this paper as systematic MDS codes with Vandermonde parity matrices serve as the base codes for \emph{bandwidth-optimal} convertible codes \cite{maturana2022bandwidth, maturana2023bandwidth} and have also been studied in various other settings \cite{lacan_fimes_2004, SHPARLINSKI2005193, roth_seroussi1985generator}.

\section{Background and Related Work}

Let us begin with an overview of important concepts and notation referred to throughout this paper, along with a literature review of previous related work.
\subsection{Systematic MDS codes and Vandermonde matrices}
\label{sec:MDS_Systematic_Vandermonde_Codes}
An $[n,k]$ linear erasure code $\Code$ with generator matrix $\V{G} \in \calM(\F)_{k \times n}$ over a finite field $\F$ is said to be systematic, or in standard form, if $\V{G} = [\V{I_k} \mid \V{P}]$ where $\V{I_k}$ is the $k \times k$ identity matrix and $\V{P}$ is a $k \times (n-k)$ matrix also known as the parity matrix. Let $\V{m}$ be a message and $\V{c}$ be its corresponding codeword under $\Code$, where $\V{m} = (m_i)_{i = 1}^k$ and $\V{c} = (c_i)_{i=1}^n$ are vectors of message and code symbols, respectively. As $\V{m}$ is encoded under $\Code$ via the multiplication $\V{c} = \V{m}^T \V{G}$, it follows that $c_i = m_i$ for all $i \le k$ if $\Code$ is systematic.

An $[n,k]$ linear erasure code $\Code$ is Maximum Distance Separable (MDS) if and only if every $k$ columns of its generator matrix $\V{G}$ are linearly independent; in other words, every $k \times k$ submatrix of $\V{G}$ is non-singular \cite{macwilliams1977theory}. As a result, data encoded by an $[n,k]$ MDS code can withstand any erasure pattern of $n-k$ out symbols in any codeword and still successfully recover the original data. If $\Code$ is also systematic with parity matrix $\V{P}$, this is equivalent to the property that every square submatrix of $\V{P}$ is non-singular \cite{macwilliams1977theory}. Such a matrix is also referred to as \emph{super-regular}. It is useful to note that any submatrix of a super-regular matrix is also super-regular.

A systematic code with a Vandermonde parity matrix $\V{P} \in \calM(\F_{k \times r})$ is one where $\V{P}$ is of the form

\begin{equation}
\label{eq:vandermonde}
\begin{bmatrix}
    1 & 1 & \dots & 1\\
    \xi_1 & \xi_2 & \dots & \xi_{r}\\
    \xi_1^2 & \xi_2^2 & \dots & \xi_{r}^2\\
    \vdots & \vdots & \ddots & \vdots \\
    \xi_1^{k-1} & \xi_2^{k-1} & \dots & \xi_{r}^{k-1}\\
\end{bmatrix}
\end{equation}

for some \emph{scalars} $\V{\xi} = (\xi_i)_{i = 1}^{r} \in \F^{r}$. Let us denote the above $k \times r$ Vandermonde matrix as $V_k(\V{\xi})$. Such a matrix is not always guaranteed to be super-regular \cite{macwilliams1977theory} and thus careful selection of the scalars is required to ensure the resulting systematic code is MDS.

\subsection{Convertible Codes \cite{maturana2022convertible}}
\label{sec:convertible_codes}

Recall that a \emph{code conversion} is a procedure that converts data from its initial representation under an $[\In, \Ik]$ code $\ICode$ to its final representation under an $[\Fn, \Fk]$ code $\FCode$. In order to capture the potential change in dimension, let $M := \mathrm{lcm}(\Ik, \Fk)$ and consider any message $\V{m} \in \F_q^M$. This is equivalent to $\Il := \frac{M}{\Ik}$ codewords in the initial configuration and $\Fl := \frac{M}{\Fk}$ codewords in the final configuration. Let $[i] := \{1, 2, \dots, i\}$ and let $|S|$ denote the size of a set $S$. Let $\V{m}[S]$ be the vector formed by projecting $\V{m}$ onto the coordinates in the set $S$, and let $\Code(\V{m})$ stand for the encoding of $\V{m}$ under the code $\Code$. Let $\Ir := \In - \Ik$ and $\Fr := \Fn - \Fk$.

\begin{definition}[Convertible Code \cite{maturana2022convertible}]
    An $(\In, \Ik; \Fn, \Fk)$ convertible code over $\F_q$ is defined by: (1) a pair of codes $(\ICode,\FCode)$ over $\F_q$ such that $\ICode$ is an $[\In, \Ik]$ code and $\FCode$ is an $[\Fn, \Fk]$ code; (2) a pair of partitions $\IPart := \{P_i^\Initial \mid i \in [\Il]\}$ and $\FPart := \{P_j^\Final \mid j \in [\Fl]\}$ of $[M = \mathrm{lcm}(\Ik, \Fk)]$ such that $|P_i^\Initial| = \Ik$ for all $P_i^\Initial \in \IPart$ and $|P_j^\Final| = \Fk$ for all $P_j^\Final \in \FPart$; and (3) a conversion procedure which, for any $m \in \F_q^M$, maps the initial set of codewords $\{\ICode(\V{m}[P_i^\Initial]) \mid P_i^\Initial \in \IPart\}$ to the corresponding set of codewords $\{\FCode(\V{m}[P_j^\Final]) \mid P_j^\Final \in \FPart\}$ over the final code.
\end{definition}

Recall that access cost during code conversion refers to the number of code symbols that are read or written during conversion. Access-optimal convertible codes are those which meet the lower bounds on access cost established in \cite{maturana2022convertible} that are known to be tight. It is known that \emph{any} $(\In, \Ik; \Fn, \Fk)$ convertible code for the merge regime where $\Ir = \Fr$ formed by a pair of systematic codes with Vandermonde parity matrices $\Ip = V_{\Ik}(\V{\xi})$ and $\Fp = V_{\Fk}(\V{\xi})$ over the same scalars is access-optimal \cite{maturana2022convertible}. This is due to $\Fp$ being ``\emph{$\Fr$-column block-constructible}'' from $\Ip$; that is, each new parity of a merged codeword can directly be computed as a linear combination of the parities of the original codewords. If the parity matrices are super-regular, then the resulting convertible code is guaranteed to be MDS as well. The best known construction \cite{maturana2022convertible} of a systematic MDS access-optimal convertible code for the merge regime where $\Ir = \Fr$ is formed by a pair of systematic codes with Vandermonde parity matrices over the scalars $\xi = (\theta^{i-1})_{i=1}^{\Ir}$, for any primitive element $\theta \in \F$. This construction requires a field size $q \ge p^D$ where $p$ is any prime and $D \in \Theta((\Fn)^3)$ \cite{maturana2022convertible}.

\subsection{Additional Notation and Preliminaries}
\label{sec:notation_prelims}

This section presents notation and terminology used in this paper that follows and expands on the notation introduced in \cite{maturana2022convertible}, and reviews some preliminaries from Galois theory that will be used in the rest of the paper.

For any two sets $I,J$, let $I \mathbin{\triangle} J$ denote the symmetric difference of $I$ and $J$. For any two integers $a,b$, let $a \perp b$ denote that $a$ and $b$ are coprime. Let $\V{x}$ denote the vector $(x_i)_{i = 1}^r$ for some $r$. Let $\V{M}_{i,j}$ denote the entry in the $i$th row and $j$th column of the matrix $\V{M}$, with both indices 1-indexed. Let $\V{M}_{I \times J}$ denote the submatrix of $\V{M}$ formed by the intersection of the rows indexed by $I$ and the columns indexed by $J$, with all indices 1-indexed. Let $\mathrm{row}_i(\V{M})$ stand for the $i$th row vector of the matrix $\V{M}$. Let $\chi_{P}$ be the indicator function for whether the proposition $P$ is true.

Let $\F_p$ denote the prime field of size $p$, and let us reserve $\F_q$ for  prime power fields of size $q = p^w$ for some prime $p$ and $w > 1$. Let $\F^\times$ denote the multiplicative group of the field, or $\F \setminus \{0\}$. Let $\mathrm{ord}(a)$ denote the order of an element $a \in \F^\times$. Let $\F[x_1,\dots,x_r]$ denote the ring of polynomials in $x_1,\dots,x_r$ over the field $\F$. Let $\mathrm{Aut}(\F)$ denote the group of automorphisms over the field $\F$. Let $S_n$ denote the group of permutations of $[n]$.

Recall that a field automorphism is a bijective map $\sigma: \F \rightarrow \F$ such that for all $x, y \in \F$, $\sigma(x+y) = \sigma(x) + \sigma(y)$ and $\sigma(xy) = \sigma(x)\sigma(y)$; in essence, the map preserves the structure of the field. Note also by definition, it must be the case that $\sigma(0) = 0$ and $\sigma(1) = 1$, which also gives us that $\sigma(-a) = -\sigma(a)$, $\sigma(a^{-1}) = \sigma(a)^{-1}$, and $\mathrm{ord}(a) = \mathrm{ord}(\sigma(a))$ for all $a \in \F^\times$. It is easy to verify that the set of fixed points of an automorphism form a sub-field of $\F$, termed the \emph{fixed field} of the automorphism. It is also a  consequence of Galois theory that the fixed field of an automorphism over the field $\F_q$ where $q = p^w$ is always an extension of the base prime field $\F_p$ \cite{dummit2003abstract}.

\subsection{Related Work}
\label{sec:related_work}

The most directly related works on access-optimal convertible codes \cite{maturana2022convertible, maturana2020access} were already discussed in \cref{sec:introduction}. In this section, we will discuss other closely related works. In addition to the access cost, previous works on convertible codes have also studied other costs of conversion such as bandwidth cost \cite{maturana2022bandwidth} and locality of repair \cite{maturana2023LRCC, Kong2023LocallyRC}. In this paper, while we focus on the access cost of conversion, the proposed new constructions do enable better constructions of bandwidth-optimal convertible codes as well. This is because access-optimal convertible codes serve as the {{base codes}} of the {{Piggybacking framework}} \cite{maturana2022bandwidth} when constructing convertible codes efficient in bandwidth cost.

There also have been previous efforts to study the fundamental limits of existence of super-regular Vandermonde matrices. Shparlinski \cite{SHPARLINSKI2005193} provided an upper bound on the total number of singular square submatrices of a Vandermonde matrix by showing that any $(q - 1) \times m$ Vandermonde matrix $V_{q-1}(\xi_1,\dots,\xi_m)$ over the field $\F_q$ has at most $3(m-1)(q-1)^mT^{\frac{-1}{m-1}}$ singular $m \times m$ square submatrices where $T := {\min}_{i \neq j \in [m]}\mathrm{ord}(\frac{\xi_i}{\xi_j})$; however, this bound has been shown to be not tight upon closer investigation \cite{lacan_fimes_2004}. Additionally, Intel's Intelligent Storage Acceleration Library (ISA-L), commonly used to implement erasure coding in practice, has published bounds on the range of parameters $[n,k]$ over $\F_{256}$ for which its code supports generation of super-regular Vandermonde parity matrices, based on a very specific construction \cite{Intel}. There is no proof provided alongside these bounds; they were likely determined by running a code script to test each submatrix for invertibility.

In addition, there has been independent work studying systematic linear MDS codes with various other constructions of super-regular parity matrices. For example, it is known that a Cauchy matrix $\V{C}$, that is, one of the form $C_{i,j} = (a_i + b_j)^{-1}$ for all $i,j \in [n]$ given two vectors $(a_i)_{i = 1}^n$ and $(b_j)_{j =1}^n$, is super-regular so long as the $a_i$'s and $b_j$'s are all distinct from each other \cite{roth_seroussi1985generator, roth_lempel1989cauchy, climent2012conv}. Additionally, Lacan and Fimes introduced a construction of super-regular matrices formed by taking the product of two Vandermonde matrices \cite{lacan_fimes_2004}. To add on, there has been considerable progress in constructing super-regular Toeplitz matrices in the development of convolutional codes \cite{Almeida2020SuperregularMO, HUTCHINSON20082585,ALMEIDA20132145}. Nonetheless, none of these alternatives are suitable for the construction of access-optimal convertible codes.

To our knowledge, in this paper we establish the best known bounds on the field size required for the existence of systematic MDS access-optimal convertible codes for the merge regime where $\Fr = \Ir$. This paper is also the first to provide, with proof, explicit constructions of systematic MDS access-optimal convertible codes for the merge regime where $\Fr = \Ir$ over practically usable field sizes.

\section{Fundamental Limits on Field Size}

\label{Sec:fundamental_limits}

In this section, we study a new construction of systematic MDS access-optimal convertible codes for the merge regime where $\Ir = \Fr$ that generalizes the construction introduced in \cite{maturana2022convertible}. The new construction is still based on systematic codes with super-regular Vandermonde parity matrices, but we allow the scalars to take on any distinct nonzero values, rather than being restricted to consecutive powers of a primitive element in the field. By virtue of the parity matrices being Vandermonde matrices, as detailed in \cref{sec:convertible_codes}, the new construction of convertible codes remains access-optimal. Thus, a proof of the existence of any $k \times r$ super-regular Vandermonde matrix yields $(\In, \Ik; \Fn, \Fk = \lam \Ik)$ systematic MDS access-optimal convertible codes for any $\lam \ge 2$, $\Fk \le k$, and $\Ir =\Fr \le r$. We will establish several bounds (\cref{thm:kr_bound,,thm:2^r_bound,,thm:k^r_bound}) on the field sizes for which there exist $[n,k]$ systematic MDS codes with Vandermonde parity matrices. This is done by studying their underlying super-regular Vandermonde matrices.

We start with a result which provides a requirement on the field sizes over which such matrices exist. This result draws upon intuition that an optimal choice of scalars for the Vandermonde matrix would avoid selecting elements with smaller order to avoid repetition along the corresponding columns.

\begin{restatable}{theorem}{thmOne}
    \label{thm:kr_bound}
    Over the field $\F_q$, a $k\times r$ super-regular Vandermonde matrix can only exist if the following condition holds: for every divisor $m$ 
    of $q-1$ where $m < k$,  $q \ge rm + 1$.
\end{restatable}
\begin{IEEEproof}
    Provided in \cref{sec:appendix_a}.
\end{IEEEproof}

For the field $\F_{256}$, for example, this result tells us that $[n=90,k=86]$ and $[n=58,k=52]$ systematic MDS codes with Vandermonde parity matrices do not exist.

The next lemma is a simple consequence of viewing finite prime power fields as vector spaces over their base prime fields. 

\begin{restatable}{lemma}{lemNonEmpty}
    \label{lem:nonempty}
    Over the field $\F_{q}$, where $q = 2^w$, for any $r > w$, for any $S = \{\xi_i\}_{i=1}^r \subseteq \F_q$, there must exist some nonempty subset $I \subseteq [r]$ such that $\sum_{i \in I} \xi_i = 0$.
\end{restatable}
\begin{IEEEproof}
    Provided in \cref{sec:appendix_a}.
\end{IEEEproof}

This lemma stems from the fact that any collection of field elements larger than the field's dimension must be linearly dependent. Over fields of characteristic $2$, this simply corresponds to a nonempty subset of elements that add to 0. This will be used later to identify a singular submatrix in a proposed Vandermonde matrix. This in turn, yields a lower bound on the minimum field size required for the existence of super-regular Vandermonde matrices specific to fields of characteristic $2$. 

\begin{restatable}{theorem}{thmTwo}
    \label{thm:2^r_bound}
   Over the field $\F_q$, where $q = 2^w$, for any $r,k$ such that $k > r$, a $k\times r$ super-regular Vandermonde matrix with distinct, nonzero scalars can only exist if $q \ge 2^r$.
\end{restatable}
\begin{IEEEproof}
    Provided in \cref{sec:appendix_a}.
\end{IEEEproof}
For example, again considering $\F_{256}$, this bound informs us that $[n=19,k=10]$ systematic MDS codes with Vandermonde parity matrices do not exist.

The first result for general fields (\cref{thm:kr_bound}) is a tighter bound for regimes where $k \gg r$ and $\exists m \approx k$ such that $m < k$ and $m$ divides $q-1$ for a proposed field size $q$; in this case, we get the bound $q^*(k,r) \ge \Omega(kr)$. On the other hand, the lower bound specific to fields of characteristic $2$ (\cref{thm:2^r_bound}) is more relevant in settings such as storage in unreliable environments which demand narrow codes with higher storage overhead, or when when $k \approx r$.

We will next prove the existence of $k \times r$ super-regular Vandermonde matrices over all fields of size greater than a threshold in terms of $k$ and $r$. We first start with a lemma that narrows down the set of square submatrices of a Vandermonde matrix that need to be tested for singularity to establish super-regularity. More specifically, we show that it is sufficient to only consider submatrices formed by a set of rows that includes the first row.

\begin{restatable}{lemma}{lemTwo}
    \label{lem:3.3}
    Over the field $\F_q$, for any $r,k,\ell$ such that $\ell \le \min(r,k)$, for any $k \times r$ Vandermonde matrix $V_k(\V{\xi})$ with $(\xi_i)_{i=1}^r \in (\F_q^\times)^r$, the submatrix $\V{H} := V_k(\V{\xi})_{I \times J}$ defined by $I := \{\alpha_1,\dots,\alpha_\ell\} \subseteq [k]$ and $J := \{\beta_1,\dots,\beta_\ell\} \subseteq [r]$, where $\alpha_i < \alpha_j$ for all $i < j$, is non-singular if and only if the submatrix $\V{H'} := V_k(\V{\xi})_{I' \times J}$ defined by $I' := \{1,\alpha_2- (\alpha_1 -1),\dots,\alpha_\ell - (\alpha_1 -1)\} \subseteq [k]$ and $J$ is non-singular.
\end{restatable}
\begin{IEEEproof}
    Provided in \cref{sec:appendix_a}.
\end{IEEEproof}

     We now utilize the Schwartz–Zippel lemma~\cite{schwartz1980,zippel1979} in a probabilistic argument for the existence of a super-regular Vandermonde matrix given a sufficiently large field size. This, in effect, establishes an \textit{upper bound} on the \emph{minimum} field size required for the existence of super-regular Vandermonde matrices.

\begin{restatable}{theorem}{thmThree}
    \label{thm:k^r_bound}
    Over the field $\F_{q}$, for any $r,k$, if $q > 1 + \binom{k}{2}\sum_{\ell = 2}^r \binom{r}{\ell}\binom{k-2}{\ell-2} \in O(k^r)$, then there must exist scalars $(\xi_i)_{i=1}^r \in (\F_q^\times)^r$ such that the $k\times r$ Vandermonde matrix $V_k(\V{\xi})$ is super-regular.
\end{restatable}
\begin{IEEEproof}
    Provided in \cref{sec:appendix_a}.
\end{IEEEproof}
Recall that the previously known upper bound \cite{maturana2022convertible} on the minimum field size $q$ required for the existence of systematic MDS access-optimal convertible codes for the merge regime where $\Ir = \Fr$ was $\log q \le \Theta((\Fn)^3)$. \cref{thm:k^r_bound} establishes the improved upper bound of $\log q \le O(\Fr \log \Fk)$, an order of magnitude smaller.

\section{Low Field Size Constructions}
\label{Sec:low_field_size_constructions}

In this section, we present several explicit constructions of systematic MDS access-optimal convertible codes in the merge regime (that is, for $(\In, \Ik; \Fn, \Fk = \lam \Ik)$ convertible codes where $\lam \ge 2$), with field sizes smaller than existing constructions. Specifically, for \textit{general prime power fields} $\F_q$ where $q = p^w$, we provide explicit constructions of convertible codes in the merge regime for all parameters such that $\Fr = \Ir \le 3$ and $w>\Fk$. For fields $\F_q$ of characteristic $2$, we present explicit constructions of convertible codes in the merge regime for all parameters such that $\Fr = \Ir \le 3$ and $q>\Fk$. We do this by providing constructions of $k \times 3$ super-regular Vandermonde matrices for field sizes: $q > p^k$ for general prime power fields (\cref{thm:general_field_construction}) and $q > k$ for finite fields of characteristic $2$ (\cref{thm:binary_field_construction}).}
These matrices serve as the parity matrices for the systematic MDS codes that underlie the aforementioned convertible codes. As every submatrix of a super-regular matrix is also super-regular, a valid parity matrix for three parities gives us one for any fewer than three parities as well.

We start with a lemma that builds on the intuition to choose primitive elements of the finite field for the scalars of the super-regular Vandermonde parity matrix.
\begin{restatable}{lemma}{lemFour}
    Over the field $\F_{q}$, for all $k < q$, given any primitive element $\theta \in \F_q$, given $2 \le e \le q-1$ such that $e, e-1 \perp q-1$, the $k \times 3$ Vandermonde matrix $V_k(1, \theta, \theta^e)$ has no singular $2 \times 2$ square submatrices.
\label{lem:4.1}
\end{restatable}
\begin{IEEEproof}
    Provided in \cref{sec:appendix_b}.
\end{IEEEproof}

Next, we introduce the idea of field automorphisms into our construction and choice of scalars, in particular as automorphisms are order preserving maps. Recall
some key properties of field automorphisms from \cref{sec:notation_prelims}.

\begin{lemma}
    Over the field $\F_{q}$ where $q = p^w$, for all $k < q$, given any primitive element $\theta \in \F_q$ and nontrivial automorphism $\sigma \in \mathrm{Aut}(\F_{q})$ with fixed field $\F_p$, the $k \times 3$ Vandermonde matrix $V_k(1, \theta, \sigma(\theta))$ has no $2 \times 2$ singular square submatrices.
\label{lem:4.2}
\end{lemma}

\begin{IEEEproof}
    First, recall that $\mathrm{Aut}(\F_{q})$ is a group generated by the Frobenius automorphism, or the map $\sigma: x \rightarrow x^p$, and thus any nontrivial element $\sigma \in \mathrm{Aut}(\F_{q})$ is of the form $\sigma(x) = x^{p^e}$ for some $1 \le e < w$. It follows that $p \le p^e < p^w = q$, and because $q \equiv 0 \mod{p}$, $q - 1 \not\equiv 0 \mod{p}$ and clearly $p^e \perp q - 1$. Next, see that if $\sigma$ has fixed field $\F_p$, this can only occur if the polynomial $p_1(x) = x^{p^e} - x$, and consequently the polynomial $p_2(x) = x^{p^e - 1} - 1$, have no roots in $\F_q$ outside of $\F_p$. This implies that $p^e - 1 \perp q - 1$, and thus we can apply \cref{lem:4.1} to get that this matrix has no $2 \times 2$ singular submatrices.
\end{IEEEproof}

For the same construction of Vandermonde matrices as in \cref{lem:4.2}, we next consider its $3 \times 3$ square submatrices and establish the necessary and sufficient conditions under which they are singular. We are able to show a significantly tighter end result for fields of characteristic $2$ in particular, but a lot of the arguments used apply to all finite fields as well. Thus, we start with an intermediate result using the shared ideas. 

\begin{restatable}{lemma}{lemFive}
    Over the field $\F_{q}$ where $q = p^w$, for all $k < q$, given any primitive element $\theta \in \F_q$ and nontrivial automorphism $\sigma \in \mathrm{Aut}(\F_{q})$ with fixed field $\F_p$, the $k \times 3$ Vandermonde matrix $V_k(1, \theta, \sigma(\theta))$ has a $3 \times 3$ singular square submatrix if and only if $\exists e_1, e_2 \in [k-1]$ and $c_1, c_2 \in \F_p^\times$ such that $e_1 < e_2$ and $\{1, \theta, \sigma(\theta)\}$ are all roots of the polynomial $f(x) = c_1 + c_2x^{e_1} + x^{e_2}$.
    \label{lem:4.3}
\end{restatable}
\begin{IEEEproof}
    Provided in \cref{sec:appendix_b}.
\end{IEEEproof}

   We now arrive at the first of our major results in this section, on explicit constructions of super-regular Vandermonde matrices over arbitrary prime power fields.

\begin{theorem}
    \label{thm:general_field_construction}
    Over the field $\F_q$ where $q = p^w$, for all $k \le w$, given any primitive element $\theta \in \F_q$ and a non-trivial automorphism $\sigma \in \mathrm{Aut}(\F_{q})$ with fixed field $\F_p$, the $k \times 3$ Vandermonde matrix $V_k(1, \theta, \sigma(\theta))$ is super-regular.
\end{theorem}

\begin{IEEEproof}
    First, note that every $1 \times 1$ submatrix of $V_k(1, \theta, \sigma(\theta))$ is non-singular as every element is a power of a nonzero element of $\F_q$. Next, by \cref{lem:4.2}, every $2 \times 2$ submatrix of $V_k(1, \theta, \sigma(\theta))$ is also non-singular. Finally, assume for sake of contradiction that $V_k(1, \theta, \sigma(\theta))$ has a singular $3 \times 3$ square submatrix. Then by \cref{lem:4.3}, $\exists e_1, e_2 \in [k-1]$ and $c_1, c_2 \in \F_p^\times$ such that $e_1 < e_2$ and $\{1, \theta, \sigma(\theta)\}$ are all roots of the polynomial $f(x) = c_1 + c_2x^{e_1} + x^{e_2}$. However, as $f \in \F_p[x]$, it must be a multiple of the minimum polynomial of $\theta$ in $\F_p[x]$, which we know is of degree $w \ge k > e_2 = \deg(f)$ as $\theta$ is a generator of $\F_q^\times$, resulting in a contradiction. Thus, every $3 \times 3$ square submatrix is also non-singular and $V_k(1, \theta, \sigma(\theta))$ is super-regular, as desired.
\end{IEEEproof}

    Finally, we show an analogous but stronger result for fields of characteristic $2$. This is of particular interest as finite fields of characteristic 2 are the most efficient choice for the representation of data in compute nodes and on storage devices.
    
\begin{theorem}
    \label{thm:binary_field_construction}
    Over the field $\F_q$ where $q = 2^w$, for all $k < q$, given any primitive element $\theta \in \F_q$ and a non-trivial automorphism $\sigma \in \mathrm{Aut}(\F_{q})$ with fixed field $\F_2$, the $k \times 3$ Vandermonde matrix $V_k(1, \theta, \sigma(\theta))$ is super-regular.
\end{theorem}
\begin{IEEEproof}
    First, note that every $1 \times 1$ submatrix of $V_k(1, \theta, \sigma(\theta))$ is non-singular as every element is a power of a nonzero element of $\F_q$. Next, by \cref{lem:4.2}, every $2 \times 2$ submatrix of $V_k(1, \theta, \sigma(\theta))$ is also non-singular. Finally, assume for sake of contradiction that $V_k(1, \theta, \sigma(\theta))$ has a singular $3 \times 3$ square submatrix. Then by \cref{lem:4.3}, $\exists e_1, e_2 \in [k-1]$ and $c_1, c_2 \in \F_2^\times$ such that $\{1, \theta, \sigma(\theta)\}$ are all roots of the polynomial $f(x) = c_1 + c_2x^{e_1} + x^{e_2}$. However, this implies $c_1 = c_2 = 1$, but then $f(1) = 1 + 1 + 1 = 1$, contradicting the fact that $1$ is a root of $f$. Therefore, every $3 \times 3$ square submatrix is also non-singular and $V_k(1, \theta, \sigma(\theta))$ is super-regular, as desired.
\end{IEEEproof}

     Using this result and the Frobenius automorphism, which is known to have fixed field $\F_p$ over any finite extension $K / \F_p$ \cite{dummit2003abstract}, we show a family of constructions of super-regular Vandermonde matrices for fields of characteristic $2$. We also give results specific to the field $\F_{256}$, which is the most commonly used finite field in practice.
    
\begin{corollary}
    Over the field $\F_q$ where $q = 2^w$, for all $k < q$, given any primitive element $\theta \in \F_q$, the $k \times 3$ Vandermonde matrix $V_k(1, \theta, \theta^2)$ is super-regular.\hfill\IEEEQEDhere
\end{corollary}
\begin{corollary}
    Over the field $\F_{256}$, for all $k < 256$, given any primitive element $\theta \in \F_{256}$, the $k \times 3$ Vandermonde matrices $V_k(1, \theta, \theta^2)$, $V_k(1, \theta, \theta^8)$, $V_k(1, \theta, \theta^{32})$, and $V_k(1, \theta, \theta^{128})$ are super-regular.\hfill\IEEEQEDhere
\end{corollary}

%%%%%%
%% To balance the columns at the last page of the paper use this
%% command:
%%
%%
%% If the balancing should occur in the middle of the references, use
%% the following trigger:
%%
\IEEEtriggeratref{10}
%%
%% which triggers a \newpage (i.e., new column) just before the given
%% reference number. Note that you need to adapt this if you modify
%% the paper.  The "triggered" command can be changed if desired:
%%
% \IEEEtriggercmd{\enlargethispage{-20cm}}
%%
%%%%%%

%%%%%%
%% References:
%%
\bibliographystyle{IEEEtran}
\bibliography{main}
%%
%% BibTeX documentation can be obtained at:
%% http://www.ctan.org/tex-archive/biblio/bibtex/contrib/doc/
%%%%%%

\newpage

%%%%%%
%% Appendix:
%% If needed a single appendix is created by
%%
%%
%% If several appendices are needed, then the command
%%
\appendices
%%
%% in combination with further \section commands can be used.
%%%%%%

\section{}
\label{sec:appendix_a}

\thmOne*
\begin{IEEEproof}
    Consider the $k\times r$ Vandermonde matrix $V_k(\V{\xi})$ for any scalars $\parens{\xi_i}_{i=1}^{r} \in \F^r_q$. Given any divisor $m$ of $q-1$, as $k > m$, the $i$th entry in the $(m+1)$th row of $V_k(\V{\xi})$ is of the form $\xi_i^m$. Note that raising this entry to the $(q-1)/m$ power would result in $\xi_i^{q-1} = 1$, so it follows that all of the entries in this row of $V_k(\V{\xi})$ are roots of the polynomial $x^{(q-1)/m} - 1$ over $\F_q$. This polynomial has exactly $(q-1)/m$ distinct roots in $\F_q$, so if $r > (q-1)/m$, then $\exists i,j \in [r]$ such that $i \neq j$ and $\xi_i^m = \xi_j^m$. It follows that $V_k(\V{\xi})_{I \times J}$ where $I = \{1, m+1\}$ and $J = \{i,j\}$ is of the form 
    $$\begin{bmatrix}
1 & 1\\
\xi_i^m & \xi_j^m
\end{bmatrix}$$
    and is singular. Therefore, in order for the matrix to be super-regular, we must have $r \le (q-1)/m \Rightarrow q \ge rm + 1$.
\end{IEEEproof}

\lemNonEmpty*
\begin{IEEEproof}
    As there are $2^r$ distinct subsets of $[r]$, and $q < 2^r$, then $\exists I,J \subseteq [r]$ such that $I \neq J$ and $\sum_{i \in I} \xi_i = \sum_{i \in J} \xi_i$. As every element is its own additive inverse in fields of characteristic $2$, it follows that $0 = \sum_{i \in I} \xi_i + \sum_{i \in J} \xi_i = \sum_{i \in I \setminus J} \xi_i + \sum_{i \in J \setminus I} \xi_i + \sum_{i \in I \cap J} \xi_i + \sum_{i \in I \cap J} \xi_i = \sum_{i \in I \mathbin{\triangle} J} \xi_i$. As $I \neq J$, $I \mathbin{\triangle} J$ must be nonempty, as desired.
\end{IEEEproof}

\thmTwo*
\begin{IEEEproof}
    Let $q < 2^r$, and consider the $k\times r$ Vandermonde matrix $V_k(\V{\xi})$ for any distinct scalars $\parens{\xi_i}_{i=1}^{r} \in (\F^\times_q)^r$ and $r,k$ such that $k > r$. Then, it follows by \cref{lem:nonempty}, that $\exists I \subseteq [r]$ nonempty such that $\sum_{i \in I} \xi_i = 0$, and we must have $|I| > 2$ as the $\xi_i$'s are nonzero and distinct. Let us define $\ell := |I|$  and $(c_i)_{i=1}^{\ell+1} \in \F^{\ell+1}_q$ to be the coefficient vector of the polynomial $f(x) := \prod_{i \in I} (x - \xi_i)$ such that $f(x) = \sum_{i = 1}^{\ell + 1} c_{i}x^{i-1}$, and note by construction $c_{\ell} = \sum_{i \in I} \xi_i = 0$. Now consider the square submatrix $\V{H} := V_k(\V{\xi})_{J \times I}$ where $J = [\ell+1] \setminus \{\ell\}$. If we take the linear combination $\V{y} = c_{\ell + 1} \mathrm{row}_{\ell}(\V{H}) + \sum_{i=1}^{\ell-1} c_{i}\mathrm{row}_i(\V{H})$, it follows that $\V{y} = (f(\xi_i))_{i \in I} = \V{0}$. As $c_{\ell+1} = 1$, this is a nontrivial linear combination of the rows of $\V{H}$, and thus $\V{H}$ is singular. Therefore, in order for the matrix to be super-regular, we must have $q \ge 2^r$.
\end{IEEEproof}

\lemTwo*
\begin{IEEEproof}
    Observe that $\V{H}$ is of the form 
$$\begin{bmatrix}
\xi^{\alpha_1 - 1}_{\beta_1} & \xi^{\alpha_1 - 1}_{\beta_2} & \dots & \xi^{\alpha_1 - 1}_{\beta_\ell}\\
\xi^{\alpha_2 - 1}_{\beta_1} & \xi^{\alpha_2- 1}_{\beta_2} & \dots & \xi^{\alpha_2 -1 }_{\beta_\ell}\\
\vdots & \vdots & \ddots & \vdots\\
\xi^{\alpha_\ell - 1}_{\beta_1} & \xi^{\alpha_\ell -1}_{\beta_2} & \dots & \xi^{\alpha_\ell -1 }_{\beta_\ell}\\
\end{bmatrix}$$
    while $\V{H'}$ is of the form
$$\begin{bmatrix}
1 & 1 & \dots & 1\\
\xi^{\alpha_2-\alpha_1}_{\beta_1} & \xi^{\alpha_2-\alpha_1}_{\beta_2} & \dots & \xi^{\alpha_2-\alpha_1}_{\beta_\ell}\\
\vdots & \vdots & \ddots & \vdots\\
\xi^{\alpha_\ell-\alpha_1}_{\beta_1} & \xi^{\alpha_\ell-\alpha_1}_{\beta_2} & \dots & \xi^{\alpha_\ell - \alpha_1}_{\beta_\ell}\\
\end{bmatrix}$$
As the $\xi_i$'s are all non-zero, it can be seen that we can get from $\V{H'}$ to $\V{H}$ by multiplying through the $i$th column by $\xi^{\alpha_1 - 1}_{\beta_i}$ for all $i \in [\ell]$. Therefore, $\det(\V{H}) = \det(\V{H'})\prod_{i= 1}^\ell \xi^{\alpha_1 - 1}_{\beta_i}$, so either $\det(\V{H}) = \det(\V{H'}) = 0$ or both matrices are non-singular, as desired.
\end{IEEEproof}

\thmThree*
\begin{IEEEproof}
    Let us start by considering an arbitrary square submatrix of our proposed $k \times r$ Vandermonde matrix $V_k(\V{\xi})$- that is, let $I := \{\alpha_1, \dots, \alpha_\ell\} \subseteq [k]$ and $J := \{\beta_1,\dots, \beta_\ell\} \subseteq [r]$ for some $\ell \le \min(k,r)$ and let us define $\V{H} := V_k(\V{\xi})_{I \times J}$ so that $\V{H}$ is an $\ell \times \ell$ submatrix of $V_k(\V{\xi})$. Observe that
    \begin{align*}
        \det(\V{H}) &= \sum_{\sigma \in S_{\ell}} \left( sgn(\sigma)\prod_{i = 1}^{\ell}  \V{H}_{i,\sigma(i)} \right)\\
        &= \sum_{\sigma \in S_{\ell}} \left( sgn(\sigma)\prod_{i = 1}^{\ell} \xi_{\beta_{\sigma(i)}}^{\alpha_i - 1} \right)
    \end{align*}
    where $S_{\ell}$ denotes the group of permutations of $[\ell]$. See that we can treat the scalars as variables $(x_i)_{i = 1}^r$ and the overall determinant as a multivariate polynomial $$f_{\V{H}}(\V{x}) = \sum_{\sigma \in S_{\ell}} \left( sgn(\sigma)\prod_{i = 1}^{\ell} x_{\beta_{\sigma(i)}}^{\alpha_i - 1} \right) \in \F_q[x_1,\dots,x_r]$$ 
    We deduce that the degree of the term in this summation corresponding to any arbitrary $\sigma \in S_{\ell}$ is $\sum_{i = 1}^\ell (\alpha_i - 1)$, and thus this is the total degree of $f_{\V{H}}$ as well. Also, note that because every term in this summation corresponds to a unique permutation of $[\ell]$ and the $\alpha_i$'s are distinct, the resulting monomial terms are also all unique, so no terms cancel out and $f_{\V{H}}$ is not identically 0 so long as $q > \deg(f_{\V{H}})$. From here, see that for any family $\calH$ of square submatrices of $V_k(\V{\xi})$, if we define $f_{\calH} := \prod_{\V{H}\in \calH} f_{\V{H}}$, then $f_{\calH}$ is also not identically 0 so long as $q > \deg(f_\calH)$. Note also that $f_{\calH}$ evaluates to 0 if and only if one of the square submatrices in $\calH$ has determinant 0 and is singular. Moreover, as $\deg(f_{\calH}) = \sum_{(I,J) \mid V_k(\V{\xi})_{I \times J} \in \calH} \sum_{\alpha_i \in I} (\alpha_i - 1)$, we can then apply Schwartz–Zippel to get that the probability that a uniformly randomly drawn vector from $(\F_q^\times)^{r}$ is a root of $f_{\calH}$ is at most 
    \begin{align*}
        \Pr_{\V{x}}\left[ f_{\calH}(\V{x}) = 0 \right] &\le \frac{\sum_{(I,J) \mid V_k(\V{\xi})_{I \times J} \in \calH} \sum_{\alpha_i \in I} (\alpha_i - 1)}{q - 1}\\
        &= \frac{\sum_{i \in [k]} (i - 1) \sum_{(I,J) \mid V_k(\V{\xi})_{I \times J} \in \calH} \chi_{i \in I}}{q-1}\\
        &= \frac{\sum_{i \in [k]} (i - 1) \left| \{V_k(\V{\xi})_{I \times J} \in \calH \mid i \in I\} \right|}{q-1}
    \end{align*}
    Now see that by \cref{lem:3.3}, it is sufficient to test for super-regularity by only considering $\calH := \{ V_k(\V{\xi})_{I \times J} \mid 1 \in I\}$. Therefore, it follows that
    \begin{align*}
        \Pr_{\V{x}}\left[ f_{\calH}(\V{x}) = 0 \right] &\le \frac{\sum_{i \in [k]} (i - 1) \left| \{V_k(\V{\xi})_{I \times J} \mid 1,i \in I\} \right|}{q-1}\\
        &=\frac{\sum_{i \in [k]} (i - 1) \sum_{\ell = 2}^r \binom{r}{\ell}\binom{k-2}{\ell-2}}{q-1}\\
        &=\frac{\binom{k}{2}\sum_{\ell = 2}^r \binom{r}{\ell}\binom{k-2}{\ell-2}}{q-1} < 1
    \end{align*}
    if $q > 1 + \binom{k}{2}\sum_{\ell = 2}^r \binom{r}{\ell}\binom{k-2}{\ell-2}$. If there is a nonzero probability that a uniformly randomly drawn vector $\V{x}$ from $(\F_q^\times)^{r}$ is not a root of any of the determinant polynomials, then there must exist some assignment of scalars $(\xi_i)_{i = 1}^r$ such that the $k \times r$ Vandermonde matrix $V_k(\V{\xi})$ is super-regular, as desired.
\end{IEEEproof}

\section{}
\label{sec:appendix_b}
\enlargethispage{-3.0cm} 
\lemFour*
\begin{IEEEproof}
    First, see that because $e \perp q - 1$, we must have that $\theta^e$ is in fact another primitive element of $\F_q$. Thus, we can handle the cases of $2 \times 2$ submatrices formed by the first and second columns or the first and third columns identically. In both of these cases, the matrix is of the form $$\begin{bmatrix}
1 & \theta^i\\
1 & \theta^j
\end{bmatrix}$$
where $q - 1 \ge k > j > i \ge 0$. It follows that the determinant of this matrix is equal to $\theta^i - \theta^j$ and thus the matrix is singular if and only if $\theta^i = \theta^j \Longleftrightarrow \theta^{j - i} = 1 \Longleftrightarrow q - 1 \mid j - i$, a contradiction as $j - i < q - 1$. Similarly, for the case that the $2 \times 2$ submatrix is formed by the second and third columns, the matrix is of the form $$\begin{bmatrix}
\theta^i & (\theta^e)^i\\
\theta^j & (\theta^e)^j
\end{bmatrix}$$
See that as $(\theta^i)^{-1}$ and $(\theta^j)^{-1}$ both exist and are nonzero, we can multiply through both rows by these constants respectively and it would not affect whether the matrix is singular or non-singular. As a result, it is sufficient to consider the matrix $$\begin{bmatrix}
1 & (\theta^{e-1})^i\\
1 & (\theta^{e-1})^j
\end{bmatrix}$$
and note that as $e - 1 \perp q - 1$, $\theta^{e-1}$ is again a primitive element, and this matrix is thus non-singular by the same proof as in the previous case.
\end{IEEEproof}

\lemFive*
\begin{IEEEproof}
    First, let us consider an arbitrary $3 \times 3$ square submatrix of the Vandermonde matrix; it must be of the form
    $$\V{H} = \begin{bmatrix}
1 & \theta^i & (\sigma(\theta))^i\\
1 & \theta^j & (\sigma(\theta))^j\\
1 & \theta^k & (\sigma(\theta))^k
\end{bmatrix}$$
    By a similar argument as in \cref{lem:4.1}, $(\theta^i)^{-1}$ and $(\sigma(\theta)^i)^{-1}$ both exist and are nonzero, as $\sigma(x) = 0$ if and only if $x = 0$. Therefore, we can multiply these values through the second and third columns, respectively, and it would not affect whether the matrix or any square submatrix is singular. Thus it suffices to consider matrices of the form
$$\V{H'} = \begin{bmatrix}
    1 & 1 & 1\\
    1 & \theta^{e_1} & (\sigma(\theta))^{e_1}\\
    1 & \theta^{e_2} & (\sigma(\theta))^{e_2}
\end{bmatrix}$$
    where $q - 1 \ge k > e_2 > e_1 > 0$. Next, see that $\V{H'}$ is singular if and only if there exists nontrivial $(c_i)_{i=1}^3 \in \F_q^3$ such that $\sum_{i= 1}^3 c_{i}\mathrm{row}_i(\V{H'}) = \V{0}$; in other words, $\{1, \theta, \sigma(\theta)\}$ are all roots of the polynomial $f(x) = c_1 + c_2x^{e_1} + c_3x^{e_2}$. See also that if $c_i = 0$ for any $i \in [3]$, then if we let $J = [3] \setminus \{i\}$, it follows that $\sum_{j \in J}c_{j}\mathrm{row}_j(\V{H'}_{[3]\times[2]}) = \V{0}$. This would equate to a singular $2 \times 2$ square submatrix of $V_k(1, \theta, \sigma(\theta))$, a direct contradiction of \cref{lem:4.2}. So we can assume the $c_i$'s are all nonzero. From here, see that $\{1, \theta, \sigma(\theta)\}$ are all roots of the polynomial $f(x) = c_1 + c_2x^{e_1} + c_3x^{e_2}$ if and only if they are also roots of the polynomial $g(x) = c_3^{-1}f(x)$, so we can assume without loss of generality that $c_3 = 1$. In summary, $\V{H'}$ is singular if and only if $\exists c_1, c_2 \in \F_q^\times$ such that $\{1, \theta, \sigma(\theta)\}$ are all roots of the polynomial $f(x) = c_1 + c_2x^{e_1} + x^{e_2}$.
    
    Plugging in our three roots into the polynomial, we get the following:
    \begin{align}
        &c_1 + c_2 + 1 = 0\label{eq:1}\\
        &c_1 + c_2\theta^{e_1} + \theta^{e_2} = 0\label{eq:2}\\
        &c_1 + c_2(\sigma(\theta))^{e_1} + (\sigma(\theta))^{e_2} = 0\label{eq:3}
    \end{align}
    Furthermore, we can plug in both sides of \cref{eq:2} into $\sigma$, giving us the additional equation
    \begin{align}
        0 = \sigma(c_1) + \sigma(c_2)(\sigma(\theta))^{e_1} + (\sigma(\theta))^{e_2}\label{eq:4}
    \end{align}
    We can combine \cref{eq:3} and \cref{eq:4} to get
    \begin{align*}
        c_1 - \sigma(c_1) = (\sigma(c_2) - c_2)(\sigma(\theta))^{e_1}
    \end{align*}
    We can then substitute $c_2 = -c_1 - 1$ from manipulating \cref{eq:1} to get
    \begin{align*}
        c_1 - \sigma(c_1) &= (c_1 + 1 + \sigma(-c_1 - 1))(\sigma(\theta))^{e_1} \\
        &= (c_1 - \sigma(c_1))(\sigma(\theta))^{e_1}
    \end{align*}
    Assume for sake of contradiction that $c_1 \notin \F_p$. Then $c_1$ is not fixed by $\sigma$ and thus $(c_1 - \sigma(c_1)) \neq 0$. We can then multiply through by $(c_1 - \sigma(c_1))^{-1}$ to get $1 = (\sigma(\theta))^{e_1}$. Note that as $\sigma(\theta)$ is a primitive element of $\F_q$, we must have $q - 1 \mid e_1$, contradicting the assumption that $e_1 < q - 1$. Therefore, we must have $c_1 \in \F_p^\times$, and as $c_2 = - (c_1 + 1)$ and fields are closed under addition and inverses, $c_2 \in \F_p^\times$ as well, as desired.
\end{IEEEproof}

\end{document}